# Transforming Student Learning with Classroom Communication Systems

Ian Beatty, University of Massachusetts Amherst
Scientific Reasoning Research Institute and
Physics Education Research Group (UMPERG)

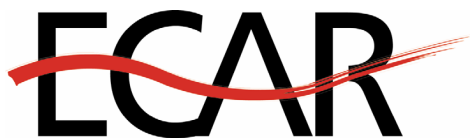

# Overview

One-on-one and in small groups, most instructors naturally ask thought-provoking questions; foster dialogue between students; encourage them to articulate and reflect on their thinking; continually probe their needs, confusions, progress, and background knowledge; and adjust teaching behavior as needed. This becomes impractical in larger classes, however, and even the most talented teachers resort to lecturing and demonstrating, perhaps asking a few rhetorical questions or calling on the occasional raised hand. The result is a mismatch between classroom practice and the instructional objectives we all claim to value; information, answers, and memory become the focus of class activity and student concern instead of conceptual understanding, process, and reasoning.[1]

Classroom communication systems (CCSs) are technology products—combinations of hardware and software—designed to support communication and interactivity in classes. Through use of these products, large "lecture" classes can function more like small discussions. In an economic context where brick-and-mortar universities face increasing competition from distance education and self-paced learning programs, they must capitalize on the fact that they bring students and faculty together face-to-face. CCSs can help them to do that.

Since 1993, the University of Massachusetts Physics Education Research Group (UMPERG) has taught with CCSs, developed curriculum and pedagogic techniques for use with them, researched CCS-based teaching, supported other CCS-adopting instructors, and interacted with CCS developers. This research bulletin will describe the nature and use of CCSs and will identify significant benefits they offer to higher education as well as challenges their use presents to instructors, administrators, support staff, and students. It will also share some advice drawn from the lessons learned through a decade of experience.

## Highlights of Classroom Communication Systems

A classroom response system is technology that

- allows an instructor to present a question or problem to the class;
- allows students to enter their answers into some kind of device; and
- instantly aggregates and summarizes students' answers for the instructor.[2,3]

A response system can conceivably be as basic as a button on every seat in the classroom and a readout dial for the instructor showing how many buttons are depressed. A CCS is a response system that provides additional support for specific student-active, question-driven, discussion-centered pedagogy, such as

- Instantly constructing a histogram of class-wide answers for the instructor

- Displaying the histogram to students via overhead projector

- Managing rosters and student logins

- Allowing an instructor to associate individual students with their answers

- Providing the instructor with a map of the classroom that displays student names and question answers by seat

- Allowing or requiring students to answer in small groups

- Supporting integrated creation, management, display, and archiving of questions

- Permitting question types other than multiple choice[4]

CCSs have the most impact in large lecture courses (50 or more students), but they can also benefit smaller classes.

## Historical Trend

Classtalk was the first popular CCS, begun in 1985 and commercially available from 1992 through 1999. It was developed by a former NASA engineer, with National Science Foundation funding and in collaboration with educational researchers at several major universities. Classtalk used common graphing calculators as student input devices (each shared by up to four students), a Macintosh computer as the instructor's command console, and a proprietary hard-wired network connecting them. It had a pedagogically sound and feature-rich design that remains the standard to which other CCS are compared. However, it was expensive, and its network required special installation in every classroom in which the system was to be used.

Starting in 1999, Classtalk was pushed out of the market by simpler, easier, and more reliable—though pedagogically more limited—response systems like EduCue PRS and eInstruction CPS. This generation of tools employs proprietary "clickers" resembling a TV remote control to send infrared (IR) signals to receptors at the front of a classroom. IR systems have achieved widespread penetration into university and K–12 classrooms.

A third generation of CCSs has now begun to appear. These systems are built from stock Internet hardware, software, and protocols. They use laptop and tablet PCs and PDAs as student input devices, Ethernet and 802.11 "wi-fi" wireless networks for connectivity, and Web browsers, HTTP, Java, and Microsoft .NET as a software base. CCSs are becoming a specialized kind of in-class Web application rather than an isolated technology. In the future, we can expect to see these systems integrate more thoroughly with other learning management and course management software. We can also expect to see them push the boundaries of the established response system model, exploring new possibilities.

## How a CCS Is Used

A CCS can be used to insert occasional audience questions into an otherwise traditional lecture, to quiz students for comprehension, or to keep them awake. These uses are a waste of the system's potential. To truly realize the benefits of a CCS, an instructor must rethink her entire instructional model and the role class time plays within it and make CCS use an integral part of an organic whole. A successful approach is to expose students to new subject material before class, perhaps through readings and Web-based multimedia. In-class time can then be devoted to CCS-mediated activities and discussions aimed at refining and extending students' understanding of the material. Following class, homework can solidify this understanding and develop related procedural skills such as quantitative problem-solving.

The following question cycle is a powerful model for organizing CCS-based teaching during in-class meetings[5,6] (see Figure 1). The instructor presents a question or problem to the class, generally without preamble, and allows the students a few minutes to discuss it among themselves in small groups. Typically, students within a group will state their various opinions and intuitions, and then offer arguments for each until one student persuades the others. Some discussion and elaboration may follow until the group is satisfied with its answer. Students then key in their answers. The instructor views an instant histogram showing the distribution of class answers and displays the histogram to the class. Without revealing which answer is correct (if any), the instructor then moderates a class-wide discussion, asking for volunteers to explain the reasoning behind each answer. With deft management, this process can be turned into a lively interchange of ideas and arguments between students. If the answers and subsequent discussion reveal a need, the instructor can follow up with a brief lecture on the relevant point.

**Figure 1: The Question Cycle—an Effective Model for CCS Use in Class**

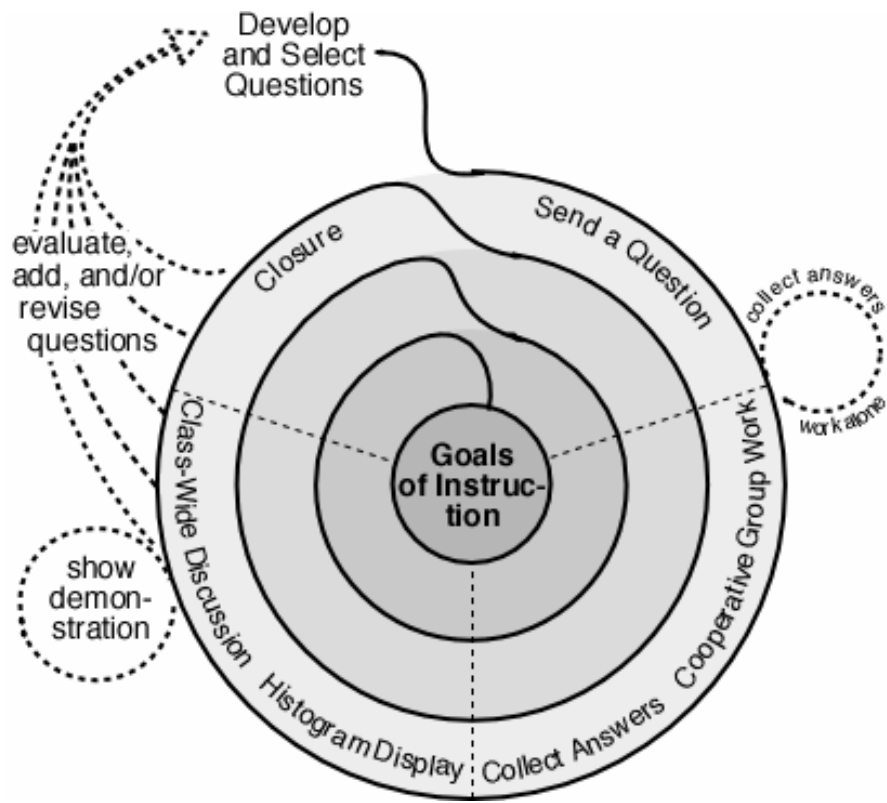

# What It Means to Higher Education

Used well, a CCS can dramatically transform the classroom environment and entire learning dynamic for a course. It can also present challenges that many instructors and universities are not accustomed to.

## Benefits

By engaging their minds in class, CCS-based instruction makes students active participants in the learning process.[5–8] This engagement results in more learning than the traditional lecture format offers and in learning of a different kind: students develop a more solid, integrated, useful understanding of concepts and their interrelationships and applicability. A concerted focus on understanding rather than recall, and on reasoning rather than answers, bolsters the effect. Merely asking rhetorical questions and pausing for students to think is insufficient; once students have committed to and externalized an answer, even if only guessing, they are emotionally invested in the problem and pay far more attention to subsequent discussion and resolution.

By providing frequent feedback to students about the limitations of their knowledge, CCS-based instruction helps them take charge of their own learning, seeking out the information and experiences they need to progress. Used consistently, it can impact their approach to learning beyond class, helping them transform into more motivated, empowered, aggressive learners. By making them conscious of their own background knowledge and preconceptions, CCS-assisted instruction can help students integrate new knowledge and overcome misconceptions.

By providing feedback to an instructor about students' background knowledge and preconceptions, CCS-based pedagogy can help the instructor design learning experiences appropriate to students' state of knowledge and explicitly confront and resolve misconceptions. By providing frequent feedback about students' ongoing learning and confusions, it can help an instructor dynamically adjust her teaching to students' real, immediate, changing needs. In the absence of such feedback, lesson plans tend to be "ballistic": they are designed and launched, and, after the fact, the instructor learns whether the target was hit. When first using a CCS, many instructors are quite shocked by how incorrect their expectations are of students' comprehension.

By having students communicate their knowledge and reasoning, in small groups and through class-wide discussion, CCS-based pedagogy can help them sharpen their vocabulary, clarify their thinking, discover gaps and contradictions in their understanding, and identify flaws in their logic. Verbalized, externalized misunderstandings are easier to dislodge, and analysis of incorrect reasoning makes correct ones easier to recognize. Students' communication and social skills also benefit. Participation in small-group discussions primes students to be more attentive to and involved in subsequent whole-class discussion. In traditional classes, students tend to ignore questions and comments by other students and only pay attention to the instructor; this tendency is reduced or eliminated in CCS-based instruction. Furthermore, students are more inclined to speak up in whole-class discussion after having first spoken in a small-group discussion. It seems that students are more afraid of being incoherent than of being incorrect.

By fostering an active, interactive classroom environment, CCS-based pedagogy helps keep students interested and attentive. CCS classes are popular with students, and they can usually articulate why. They appreciate the system's value for engaging them in the material. They acknowledge that hearing other students' reasoning helps to clarify their own. They particularly like class-wide histograms: they like the reassurance that they're not alone even when they're wrong, as well as the perception that they're part of a "community of learners" all struggling with the same ideas. They also find CCS classes more entertaining. I have never seen a student doze off during a CCS-based class.

## Challenges

**New Roles for Instructors.** Perhaps the largest barrier to the adoption of CCS-based teaching is the fact that instructors must learn new skills and adjust to new roles, and this can be intimidating and demanding. Obviously, an instructor must master the technical skills required by the particular CCS chosen: authoring, editing, and arranging questions; controlling the system in class and interpreting the data it provides; and troubleshooting technical glitches. More fundamentally, an instructor must learn to think of herself as an engineer of learning experiences rather than as a dispenser of knowledge. She must learn to plan curriculum around questions and deep comprehension, rather than around lecture notes and content coverage. The art of designing effective questions is deceptively nontrivial and can be time-consuming for an instructor new to CCSs. (With experience, the process can become as efficient as traditional lecture planning.)

An instructor must also learn the art of class management—soliciting and moderating discussion and directing students' attention. Using information revealed through CCS answers and discussion to model students' mental state, and responding appropriately, requires quick thinking and skills that few instructors have developed. Student-active pedagogy also requires the instructor to assume the role of "learning coach," meta-communicating about the learning process and students' approach to it.

The most daunting aspect for many instructors may be the necessity of giving up control of the class. A lecture is predictable and controlled, with attention safely focused on the instructor. CCS-based teaching, on the other hand, necessarily turns the classroom over to students while they debate in small groups and while they discuss their reasoning after the histogram display. An instructor must learn to steer the apparent chaos, rather than rigidly controlling or squelching it. Furthermore, if an instructor is serious about using formative assessment to respond to the students' needs as they are revealed, she must be prepared to adjust or abandon any existing lesson plan and extemporize.

The best way to help instructors adjust to their new roles is to provide mentoring and support by CCS-experienced teachers. A little scaffolding can go a long way. Adaptability and a willingness to improvise, experiment, and learn from apparent failures are important attitudes to encourage. Sharing questions between instructors, or even providing a library or model curriculum of predesigned question sets, can make a big difference to a new instructor trying to climb the steep CCS learning curve.

**New Roles for Students.** Although students generally express positive feelings about CCS-based instruction after they have adjusted to it, some initially greet it with fear and discomfort. This reaction is most prevalent among students accustomed to doing well: they have mastered the game, and now the rules are being changed. Others are resentful out of simple laziness: they are being asked to engage in thought and activity during class which, though beneficial, is effortful and at times frustrating. Many are uncomfortable with the idea that they will be held accountable for material not directly addressed in lecture. Also, they must prepare beforehand so they will be ready to participate during class.

Students often perceive CCS-delivered questions as mini-tests that they "ought" to be able to answer correctly. If the questions address novel contexts, subtleties, or ambiguities (as good questions should), they may be perceived as unfair. This is because students are accustomed to summative assessment. They don't recognize that they are supposed to wrestle with the questions and that true learning occurs in the struggle and resolution.

The best way to help students adjust to their new roles is to meta-communicate about the purpose and design of the pedagogy, enlisting students as collaborators in their own education. Consistently directing their attention toward learning rather than evaluation (grades), and toward reasoning rather than answers, is important.

**New Roles for Administrators and Support Staff.** If an institution wishes to support the use of CCS-based pedagogy by its instructors, it must provide more pre-class and in-class technical support than conventional lecturing requires. At some universities, CCS support can get lost in the cracks, falling within neither the customary purview of traditional classroom technology support (handled by an audio-visual department) nor traditional computer and network support (handled by central IT). High-level policy directives may be required.

Technical support is not sufficient, however. Instructors, especially those just starting out with a CCS, need instructional support and mentoring as recommended above. If an institution is committed to supporting CCS use, it should consider creating an institute, center, program, or other formal structure with this explicit charge.

Some classrooms are physically more CCS-friendly than others. Having one or more large projection screens at the front of the room, easily readable from all seats, is required. A room's seating arrangement must allow for convenient small-group discussions. Depending on the size and acoustics of the room, mobile microphones and a sound system may be indicated. If a laptop-based system is chosen, table space for laptops is important, wireless or wired network access is crucial, and electrical outlets throughout the room are valuable. Schools may wish to renovate some classrooms specifically to support CCS use.

**Incompatibility with Old Metrics.** Changing one component of a tightly coupled system is generally unsuccessful. CCS-based pedagogy stresses rich conceptual understanding, reasoning, and knowledge transfer, and it will fail if students are assessed by traditional metrics focusing on content recall and answers. Appropriate assessment is possible even via multiple-choice, machine-graded tests, but exam questions must be constructed with the same thoughtfulness as CCS questions.

In addition, instructors using a CCS will encounter conflict with existing notions of how much subject material ought to be covered in a course, and at what rate. CCS-based courses rarely cover as long a list of topics as traditional courses. This is not to say that students are learning less; few in a traditional lecture course grasp everything mentioned by an instructor, and much of what they do grasp is lost shortly after the last exam is taken. CCS-based pedagogy tends to be initially slow as students become accustomed to the technology and the teaching model. Also, instructors and students are taking time to build and explore a solid conceptual edifice rather than shoveling content and moving on. The efficiency of CCS pedagogy depends on whether one is measuring what students learn or what they've been exposed to.

## Advice

I conclude with some advice based on experiences with CCS teaching and mentoring.

**Use Appropriate Pedagogy.** Technology doesn't inherently improve learning; it merely makes possible more effective pedagogy, and only when it is consonant with an instructor's educational philosophy and beliefs and reinforced by other components of the total course. An instructor cannot and should not explicitly address in class every topic, idea, fact, term, and procedure for which students are "responsible." Instead, use class time to build a solid understanding of core concepts, and let pre-class reading and post-class homework provide the rest. Use exams and other performance metrics that support, not contradict, the concept- and reasoning-centered focus of the class.

Effective CCS use requires us to design for "multi-pass learning," in which an idea or technique is developed through multiple visitations in varying contexts spread over time. We must also appreciate, and convince our students, that confusion is an inevitable part of learning and that no lecture or lesson plan is so perfect that students will fully understand what is taught at first encounter.

If a CCS-using class is to be more than an endless series of quizzes, we should focus on the reasoning behind answers and not on their correctness, and students must be convinced that the questions are for learning and not for evaluation. How we respond when right or wrong answers are offered is crucial. A full spectrum of answers should be drawn out and discussed before we give any indication which (if any) is correct. Even the notion of "incorrect" should be downplayed; it is more enlightening to students, and more conducive to discussion, to say something like "that would be correct if…," identifying the circumstances or assumptions under which the answer would be right. Instead of offering the wrong answer, students often offer the right answer to the wrong question.

**Avoid the "Instructor-Centric" Classroom.** As teachers, we are accustomed to being the focus of attention in our classrooms. To use a CCS effectively, we must learn to give up control and allow learning to occur without constant micromanagement. When we present a question, we should resist the temptation to read the question out loud or clarify it. If it contains ambiguities, it's better to allow class discussion and student questions to bring them out; we should learn to be quiet and wait while students read, discuss, and answer a question. During class-wide discussion, we should be tolerant of silences while students ponder and make up their minds to speak out. It is at times appropriate to paraphrase a student's statements for the rest of the class or to help a struggling student express an idea, but we should always confirm our accuracy with the originator. Rather than jumping on errors or flaws in an argument, we should whenever possible allow other students to find them, even if this appears inefficient.

Empirically, two to four questions per 50-minute class is optimal. If we cannot fill a class with that many, it probably means we aren't sufficiently cultivating discussion or posing sufficiently divisive questions. When deciding how long to allow for small-group discussion, sound can be a clue: the noise level in the room tends to rise as students finish reading and assimilating the question and begin discussing it, and then drops as they reach resolution and enter their answers. If too much more time passes, it rises again as small talk ensues.

**Use Question Wrap-Up.** The transition period that wraps up class-wide discussion of one question and either transitions to another question or ends class is an important opportunity. We can summarize the key points or arguments students have put forth, possibly adding additional ones that students missed. We can also make connections to related questions and topics, pose "what if" alternative questions to be pondered but not answered, or segue into the next question. Before we've revealed which answer is "right" (or which answers are right), we may ask for a show of hands indicating how many students have changed their minds as a result of the discussion. If the count is significant, we can resend the question to see how the histogram differs.

If student answers and discussion have revealed a fundamental gap in knowledge or understanding that discussion did not resolve, a mini-lecture on some piece of subject matter may be appropriate. Students will learn far more from it than they would without the preceding CCS question because the presentation is now motivated and contextualized. Alternatively, the instructor may help students structure their knowledge and avoid getting lost in details by communicating about how material just covered fits into the larger picture of the subject as a whole or of learning in general.

Although this phase of the question cycle is necessarily instructor centered, take care to keep it relatively short and directly tied to students' recent and upcoming learning activities and do not slip into extended lecturing.

**Engineer Questions Deliberately.** In CCS-based pedagogy, question design replaces lecture note preparation as the focus of class planning. The criteria for an effective CCS question are quite different from those for exam, quiz, or homework questions, and question design should be given great care. Each instructor must discover what kinds of questions suit his or her own subject and style; however, a few broad principles apply.

Every question should have a clearly identifiable pedagogic goal, not just a topic to address. The goal indicates the action we hope to induce in students' minds. Some general types of goals include

- Drawing out students' background knowledge and beliefs on a topic

- Making students aware of their own and others' perceptions of a situation

- Discovering points of confusion or misconception

- Distinguishing two related concepts

- Realizing parallels or connections between different ideas

- Elaborating the understanding of a concept

- Exploring the implications of an idea in a new or extended context

In general, avoid computational or simple factual questions or those that probe memory rather than understanding. Comparison questions are powerful, as are predictions and causal relationships (for example, "What would happen to X if Y were increased?"). Strive for questions that get students to reason qualitatively and to draw conclusions from a conceptual model. If the instructor can anticipate likely misunderstandings and points of confusion, she should design questions to "catch" students in those and thus draw them out for discussion and resolution.

Ambiguity is a good quality of a CCS question. It sensitizes students to the ambiguous point's implications, trains them to pay attention to subtleties in a situation, and motivates a discussion about what aspects of a question statement are important and how they matter. All this may not help students reach a correct answer to the question at hand, but the answer isn't the goal: it helps students learn to reason and think defensively and to answer future questions, especially the vague, fuzzy kind often encountered outside the classroom.

A broadly spread histogram of student answers, indicating several popular choices, is a signature of an effective question. It provides good material for discussion and argument among students and is likely to result in significant learning all around. Conversely, a histogram with a single peak at the right answer accomplishes little except to indicate that students can answer that particular question and that it warrants no further time.

Sequences of related questions can build on each other to develop a complex idea or set of related ideas. One tactic is to present a concept in different contexts, helping students separate the details of application from the concept's essence. Another is to use slight variations of the same question—perhaps different questions about the same situation—to explore the limitations of a concept or to relate different concepts. In general, use familiar situations for new concepts to develop understanding; use new situations for familiar concepts to check for understanding (as on an exam). This differs from common practice. Keep in mind the issue of "cognitive load": it can take significant mental resources for students to process and interpret a new problem situation or description, and once students have done so, it is efficient to reuse that situation for multiple questions.

Occasionally, including extra information in question statements or omitting necessary information is also beneficial. It helps students learn to decide for themselves what information they need to answer a question, a skill that is perhaps as valuable in real life as determining the answer itself.

Finally, when building a lesson, when and how a question is presented impacts the depth, quality, and nature of the resulting group work and student thought. Students will naturally assume that the question is relevant to whatever has just transpired, and this can lead to "pigeonhole" learning in which concepts are not structured in a broadly useful hierarchy, but are learned chronologically and only accessible within a narrow context. In general, if a question is posed before presentation or coverage of subject material, students will draw on preexisting knowledge, apply intuition, and extrapolate from prior course material. If it is presented after, they will draw on whatever was just covered regardless of its relevance. Which is preferable depends on our pedagogic goals for the question.

**Meta-communicate.** The most powerful tool for changing students' attitudes about learning and enlisting them as active collaborators in their own education is meta-communication—high-level communication about the nature and purpose of the "normal" communication within the course. Meta-communication can and should address the learning objectives of the course and its components, the virtues of instructional techniques and styles employed, and the reasons why particular assignments are given. Experience shows that students are far more cooperative when they understand why they are being asked to do something.

Meta-communication can also address individual learning habits and the dynamics of small-group and whole-class discussion. By accepting the role of a learning coach and advisor, we can help students find strategies that increase their benefit from the course and from education in general—that is, help them learn how to learn. This is particularly important with students who display frustration or hostility toward CCS use and associated pedagogy.

## Key Questions to Ask

- Are CCS use and its associated pedagogic perspective and methods consistent with your institutional mission? Should you actively encourage it?

- Will students encounter CCS-style pedagogy and learning expectations consistently across courses or only in isolation?

- Does your institution's larger structure of exams and course interdependencies, with the curriculum content expectations they imply, support or conflict with the more thorough and less topically broad nature of CCS-based instruction?

- Are instructors expected to master technical aspects of CCS use themselves, create their own content, and refine their own pedagogy? Can institution-level support be provided?

- Will the fruits of one instructor's work developing curriculum and pedagogy for a CCS-based course be preserved if that instructor leaves the course?

- Is a candidate CCS pedagogically rich and well-designed? Do its creators have educational experience and credentials? Will it integrate well with other educational software, now and in the future? Is it extensible?

## Where to Learn More

- UMass Physics Education Research Group's Assessing Student Knowledge with Instructional Technology (ASK-IT) project's Web site provides more information and links to relevant papers. It also maintains an extensive list of available systems, with links to manufacturers' Web sites, <http://umperg.physics.umass.edu/projects/ASKIT>.

- The Assessing to Learn (A2L) Web site (another UMPERG project) has a database of CCS-compatible formative assessment questions designed for high school physics instruction and some supporting material, <http://A2L.physics.umass.edu/>.

- E. Mazur's *Peer Instruction: A User's Manual* (Upper Saddle River, N.J.: Prentice Hall, 1997) presents a detailed account of one instructor's use of CCS-based pedagogy, with extensive question sets and other resources for physics teachers.

- D. Johnson, R. Johnson, and K. Smith's *Active Learning: Cooperation in the College Classroom* (Edina, Minn.: Interaction Book Co., 1991) contains an introduction to, background on, and resources for using cooperative learning in college instruction.

- The Active Learning Web site is a compendium of resources with a bibliography for those interested in "active learning," <http://www.active-learning-site.com/>.

## Endnotes

## About the Author

*Ian Beatty (beatty@physics.umass.edu) is a postdoctoral research associate with the University of Massachusetts Amherst Scientific Reasoning Research Institute and Physics Education Research Group (UMPERG).*

ECAR